\documentclass[%
preprint,
nobibnotes,
 amsmath,amssymb,
 aps,
prb,
]{revtex4-2}

\usepackage{amsmath}
\usepackage{hyperref}
\usepackage{graphicx}
\usepackage{epstopdf}
\usepackage{bm}
\usepackage{color}
\usepackage{diagbox} 
\usepackage{dcolumn}

\begin{document}



\title{
Extreme Nanoconfinement Reshapes the Self-Dissociation of Water
}

\author{Chenyu Wang}%
\author{Wanjian Yin}%
\author{Ke Zhou}%
\email{zhouke@suda.edu.cn}
\affiliation
{College of Energy, SIEMIS, Soochow University, Suzhou 215006, China}

%

\date{\today}

\begin{abstract}


Water’s ability to self-dissociate into H$_3$O$^+$ and OH$^-$ ions is central to acid–base chemistry and bioenergetics. 
Recent experimental advances have enabled the confinement of water down to the nanometre scale, even to the single-molecule limit, yet how this process is altered at the extreme nanoconfinement remains unclear. 
Using \emph{ab-initio} calculations and enhanced-sampling machine-learning potential molecular dynamics, we show that monolayer-confined water exhibits a markedly lower barrier to auto-dissociation than bulk water.  
Confinement restructures both intramolecular bonding and the intermolecular hydrogen-bond network, while enforcing quasi-2D dipolar correlations that amplify dielectric fluctuations.
Our results imply that two-dimensional confined water could act as a \emph{superdielectric} medium and may exhibit \emph{superionic} behavior, as observed in recent experiments. 
These findings reveal confinement as a powerful route to enhanced proton activity, shedding light on geochemical niches, biomolecular environments, and nanofluidic systems where water’s chemistry is fundamentally reshaped.

\end{abstract}

\maketitle
\newpage 
 

\section{Introduction}

Water is the most abundant and indispensable solvent in nature, with its remarkable properties originating from its polar molecular structure and collective hydrogen-bond (HB) network \cite{eisenberg2005_water,marcus1985}.
A striking consequence is its unusually large dielectric constant ($\varepsilon_{\rm bulk} \approx 80$), which effectively screens electrostatic interactions, stabilizes charged species, and reduces the energetic cost of ion solvation \cite{eisenberg2005_water,marcus1985,munoz2021_chemrev}. 
This exceptional dielectric response underpins water’s extraordinary dissolving power, making it the indispensable medium for chemistry, biology, and technology.
Among its many functions, the spontaneous self-dissociation of water,
${\rm{2H_{2}O}} \rightleftharpoons  \rm{H_3O^{+}+OH^{-}}$,
is of central importance, which defines the pH scale, governs acid–base chemistry, and  provides the foundation for proton generation in biological and technological systems \cite{eisenberg2005_water,marcus1985,munoz2021_chemrev,liu2023_prl,munoz2017_prl}. 
The process of self-dissociation is a rare event, which is controlled by the dynamical nature of OH covalent bonds, collective HB rearrangements and dipolar fluctuations, with a free energy barrier ($\Delta F$) of $\approx 15$ kcal/mol$^{-1}$ \cite{liu2023_prl,munoz2017_prl,joutsuka2022_jpcb}.
Empirically, the value of $\Delta F$ is correlated with the $\varepsilon$, with higher $\varepsilon$ leading to lower $\Delta F$ for liquid water (see experimental results in Fig. S1) \cite{munoz2017_prl}. 
In bulk water, the free-energy barrier for self-dissociation has been extensively studied, with dielectric screening recognized as a key factor in stabilizing the separated ions \cite{liu2023_prl,munoz2017_prl,calegari2023_pnas,grifoni2019_pnas}. 
These suggest that any factor capable of modifying water’s HB network or dielectric response, such as spatial confinement, could significantly influence its self-dissociation.

Recent experimental advances have enabled the confinement of water down to the nanometre scale, and even to the single-molecule limit \cite{geim2018_science,geim2017_science,xue2021_science,cheng2018_natnano}. 
Such conditions are in fact ubiquitous in geochemical pores, biomolecular environments, and nanofluidic devices. 
Nanoconfinement profoundly reshapes the HB network, thereby altering the dynamics and dielectric response of water \cite{geim2018_science}. 
This leads to molecular layering, orientational polarization, disrupted hydrogen bonding, and modified dielectric properties \cite{munoz2021_chemrev,trushin2025_nrp}. 
These effects give rise to extraordinary phenomena such as rich polymorphism, melting-point shifts, pronounced dielectric anisotropy, violations of the ice rules, and even room-temperature superionic states \cite{trushin2025_nrp,kapil2022_nature,strano2017_natnano,
geim2018_science,jiang2024_natphy}.
Despite these advances, the thermodynamic behavior of water auto-dissociation under extreme confinement, particularly in the monolayer limit, remains poorly understood. 
Previous studies of confined water have reported conflicting results for self-dissociation,  including enhancement, suppression, or even insensitivity of the barriers compared to bulk \cite{munoz2017_prl,sirkin2018_jpcl,di2023_angew,dasgupta2025_jacs,advincula2025_arxiv}.
Resolving this controversy requires mechanistic insight into how confinement reshapes the bonding nature and collective dynamics of HB network.

Recent experimental studies indicate that both the conductivity and the in-plane dielectric constant of water ($\varepsilon_{\parallel} \sim 100$–$1000$) increase dramatically as the thickness approaches the molecular scale \cite{wang2024_arxiv}. 
This behavior aligns well with the characteristics of superionic liquids and reminiscent of the superionic water observed in extremely confined water at room temperature in contrast to bulk superionic water in extremely high temperature and pressure \cite{kapil2022_nature,jiang2024_natphy,coles2025_arxiv}.
These results directly indicate that nanoconfinement lowers the free-energy barrier $\Delta F$ for water self-dissociation \cite{munoz2017_prl}, yet mechanistic insight into how confinement reshapes the thermodynamics of ionization remains scarce. 
Here, we address this issue using machine-learning molecular dynamics (MLP-MD) to investigate water self-dissociation in both bulk and monolayer (1L) confinement. We find that $\Delta F$ is significantly lower in 1L water compared to bulk.
Analysis of maximally localized Wannier functions (MLWFs), vibrational density of states, and HB structure reveals that geometric confinement induces frustrated HBs, which in turn facilitate OH covalent bond cleavage and the separation/hydration of H$_3$O$^+$ and OH$^-$. 
Remarkably, we uncover a brand new dissociation pathway associated with disrupted HBs and interstitial, non-hydrogen-bonded water molecules with 55$^\circ$ motif.
This reduction in $\Delta F$ implies the anomalously enhanced $\varepsilon_{\parallel}$, which promotes charge separation and stabilizes the dissociated ions.
The picture is further supported by the results of \emph{ab-initio} MD simulations for ion hydration and ion pairing. Moreover, we find that higher water density and nuclear quantum effects further facilitate the autoionization process.
Altogether, our findings establish a direct link between dielectric anisotropy and chemical reactivity in confined water, providing a new framework for understanding and ultimately controlling ionization processes at the nanoscale. They further reveal that confined water acts as an active mediator of chemical reactions rather than a passive solvent \cite{li2025_natchem}.

\section{Results}

\begin{figure*}[tbhp]
\centering
\includegraphics[width=1.0\textwidth]{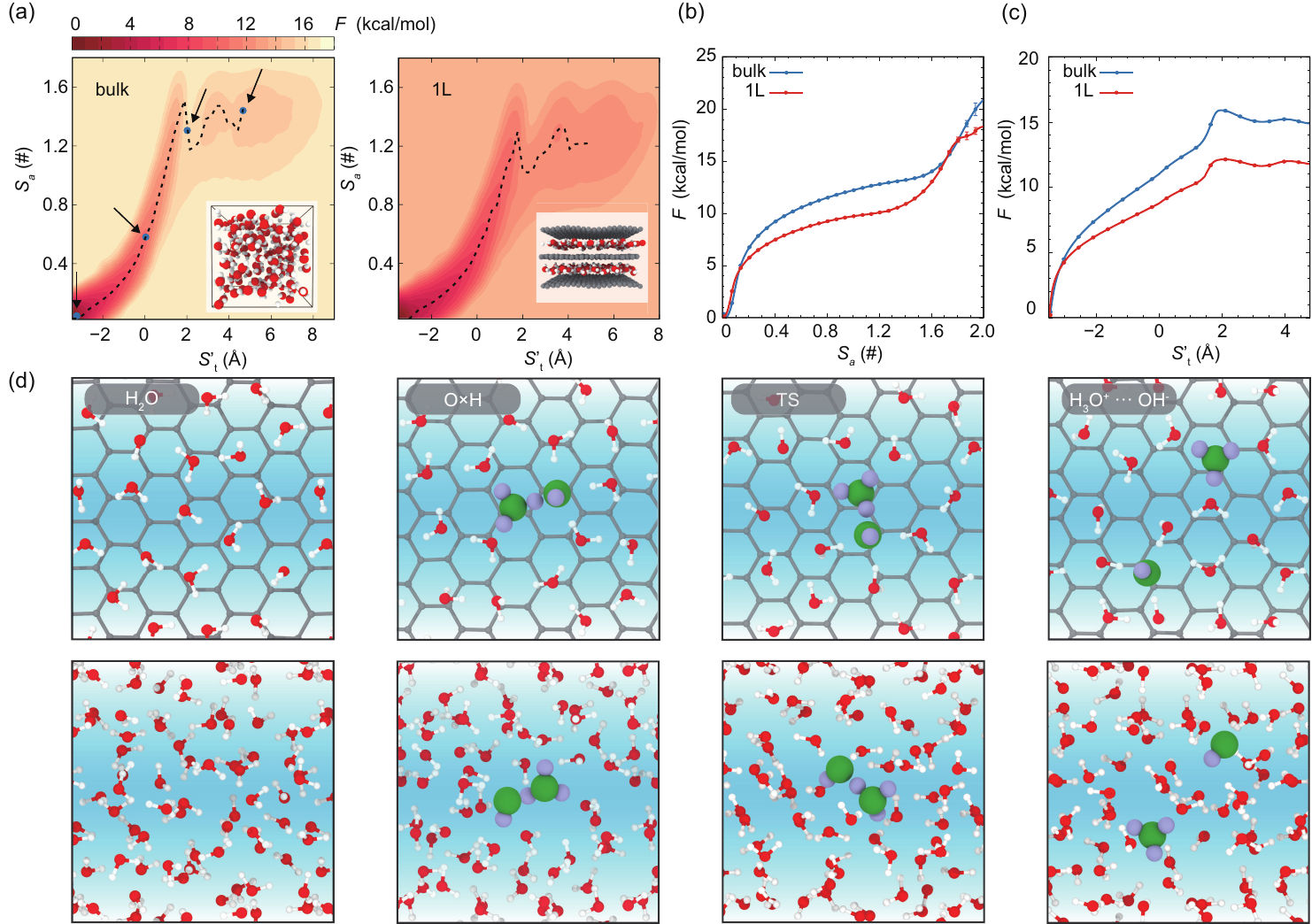} 
\caption{
(a) Two-dimensional free energy surface (FES) of water self-dissociation for bulk (left panel) and 1L (right panel) with the CVs of $S_a$ and $S_t’$ (see exact definitions in \textbf{Methods}).
The representative states are indicated by arrows.
The inset is the illustration of the simulation models.
(b) The one-dimensional FES along $S_a$ and (c) $S_t’$. 
(d) The snapshots for representative states for water self-dissociation in 1L (Top row) and bulk (Bottom row) water. 
Here ``H$_2$O'' represents pure water with no ion.
The gray, red, white balls are C, O and H atom, respectively.
``$\text{O} \times \text{H}$'' is the state with the breaking of the OH covalent bonds, 
``TS'' is the transition state that the H$_2$O has just dissociated into H$_3$O$^+$ and OH$^-$ forming a contact ion pair,
``H$_3$O$^+$···OH$^-$'' is the finally state the water molecule is fully dissociated where H$_3$O$^+$ and OH$^-$ are separated.
The atoms for H$_3$O$^+$ and OH$^-$ are highlighted by large atoms.
} 
\label{fig1}
\end{figure*}

\textbf{MLP simulation of water self-dissociation.}
Because the free-energy barrier $\Delta F$ is much larger than $k_{\rm B}T$ ($\Delta F \gg k_{\rm B}T$), the probability of spontaneous auto-dissociation is extremely low. Therefore, enhanced-sampling methods are required to explore this process. Here, we employ two advanced collective variables (CVs) based on the Voronoi space partition method (see \textbf{Methods}) \cite{grifoni2019_pnas,liu2023_prl,zhang2025_Langmuir}. The first CV quantifies the number of auto-dissociated ions ($S_a$), while the second measures the distance between H$_3$O$^+$ and OH$^-$ ($S_t’$).
In many previous studies, the coordination number of a selected oxygen atom (O$^*$) with surrounding H atoms ($n_{\rm H}$) has been used as the CV to describe dissociation. However, this simple descriptor lacks a well-defined product and transition state, making the calculated $\Delta F$ ambiguous \cite{grifoni2019_pnas,liu2023_prl,zhang2025_Langmuir,joutsuka2022_jpcb}. Moreover, this approach implicitly assumes that dissociation occurs only at the chosen O$^*$ (local), whereas in reality, it can occur at any water molecule in the system (global). By contrast, our advanced CVs overcome these limitations and provide a robust description of the auto-dissociation process globally.
Here, we employ the RPBE+D3 functional to generate the DFT data used for training the MLP.
Unlike some GGA and meta-GGA functionals, which suffer from artificial high-temperature  (AHT) behavior and overstructuring \cite{zhou2024_jctca,morawietz2016_pnas}, RPBE+D3 provides an accurate description of liquid water structure and the dynamical nature of the HB network, approaching the accuracy of CCSD(T) (see details in Figs. S2-S3).
Notably, this performance is achieved despite RPBE+D3 is GGA-level functional on Jacob’s ladder of DFT approximation.
In this work, we mainly focus on the results of 1L and bulk water. Some results of ice I\emph{h} are also included for comparison.

\begin{table*}[htbp] 
\centering
\setlength{\tabcolsep}{5.0mm}
\renewcommand{\arraystretch}{0.800}
\caption{
Table summary of the calculated free energy barriers for water auto-dissociation ($\Delta F$) and pKw using various methods.
The experimental data is also listed for comparison. 
The results in the brackets are calculated by PIMD simulations.
}
\bigskip
\begin{tabular}{cccc}
\hline
Method & $T$ (K) & $\Delta F$ (kcal/mol) &  pKw  
\\
\hline
DC-$r^2$SCAN \cite{dasgupta2025_jpcl}    & 300   &  14.9 (11.0) & 13.7  \\
SCAN \cite{calegari2023_pnas}            & 300   &  14.5   & 14.7      \\
M06-2X \cite{zhang2025_Langmuir}         & 300   &  20.8    &  15.21          \\
revPBE+D3 \cite{joutsuka2022_jpcb}       & 298   &  18.7    & 13.7     \\
RPBE+D3 \cite{liu2023_prl}               & 298   &  18.9 (11.9) & 14.1     \\
BLYP \cite{galli1993_cp}                 & 300   &  17.4    &   12.7         \\
this work, bulk		                     & 300   &  14.9 (10.5) &  14.8   \\
this work, 1L		                     & 300   &  11.8 (8.1)  &        \\
\hline
Exp., \cite{1955_exp}  & 298   &  16.7  &  15.7      \\
Exp., I\emph{h} ice \cite{eisenberg2005_water}                      & 263   &  25.7 &      19.7      \\
\hline
\end{tabular}
\label{table1}
\end{table*}

\textbf{Enhanced self-dissociation for 1L Water.}
The calculated two-dimensional free-energy surface (FES), $\Delta F(S_a, S_t’)$, for water self-dissociation is shown in Fig. \ref{fig1}a.
An apparent dissociation pathway emerges (see snapshots in Fig. \ref{fig1}d), starting from the neutral water state ($S_a \approx 0$, $S_t’ \approx -3.5$ {\AA}), progressing to the breaking of the OH covalent bond ($S_a \approx 0.7$, $S_t’ \approx 0.2$ {\AA}, $\text{O} \times \text{H}$, see details in \textbf{Methods}), then reaching the transition state (TS, $S_a \approx 1.3$, $S_t’ \approx 2.0$ {\AA}), where H$_2$O has just dissociated into H$_3$O$^+$ and OH$^-$ forming a contact ion pair, and finally to the fully dissociated state where H$_3$O$^+$ and OH$^-$ are separated (H$_3$O$^+$···OH$^-$, $S_t’ \approx 4.8$ {\AA}, $S_t \approx 6.0$ {\AA}).
The one-dimensional FES along $S_a$ and $S_t’$ was obtained by reweighting $\Delta F(S_a, S_t’)$ (Fig. \ref{fig1}b). These results clearly show that $\Delta F$ for 1L water is lower than that for bulk water. In particular, the barrier extracted from the 1D FES of $\Delta F(S_t’)$ (corresponding to the level of H$_3$O$^+$···OH$^-$ state) is 11.8 kcal/mol for 1L water and 14.9 kcal/mol for bulk. This implies that the probability of observing self-dissociation in 1L water is $\exp (3.11{\rm kcal/mol}/RT) \approx 200$ times higher than in bulk.
Further path-integral molecular dynamics (PIMD) simulations show that nuclear quantum effects (NQEs) facilitate dissociation in both cases, while the barrier for 1L water remains consistently lower than for bulk (Fig. S4). 
Notably, our calculated values of $\Delta F$ and pKw for bulk water agree well with previously reported simulation results and experimental data (Table \ref{table1}).
The discrepancy between simulation results may from the choose of CVs and exchange-correlation functionals.

\begin{figure}[tbhp]
\centering
\includegraphics[width=0.40\textwidth]{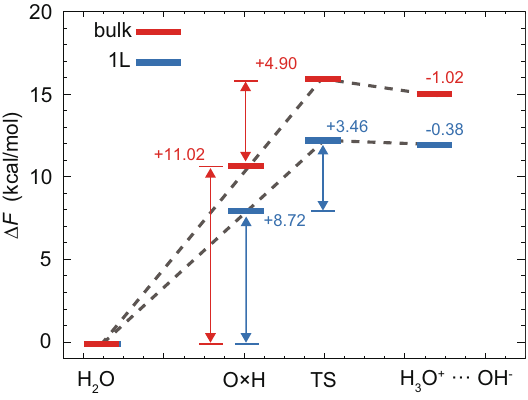} 
\caption{
The free energy level for representative states for bulk water and 1L water. 
} 
\label{fig2}
\end{figure}

\textbf{Reshapes water self-dissociation under extremely confinement.}
The free-energy levels of the four states are shown in Fig. \ref{fig2}. 
The results indicate that the reduction of $\Delta F$ is mainly determined by the first and second steps. 
The first step corresponds to the cleavage of the O–H covalent bond, which is directly related to bond strength and the electronegativity of the oxygen atom. 
The second step involves the separation of the transient $\text{O} \times \text{H}$ state into well-defined H$_3$O$^+$ and OH$^-$ ion pairs, which depends strongly on the stabilization of charges by the local dielectric environment. 
Together, these processes reflect both the electronic structure of the O–H bond and the dynamical nature HB network.

\begin{figure*}[tbhp]
\centering
\includegraphics[width=0.80\textwidth]{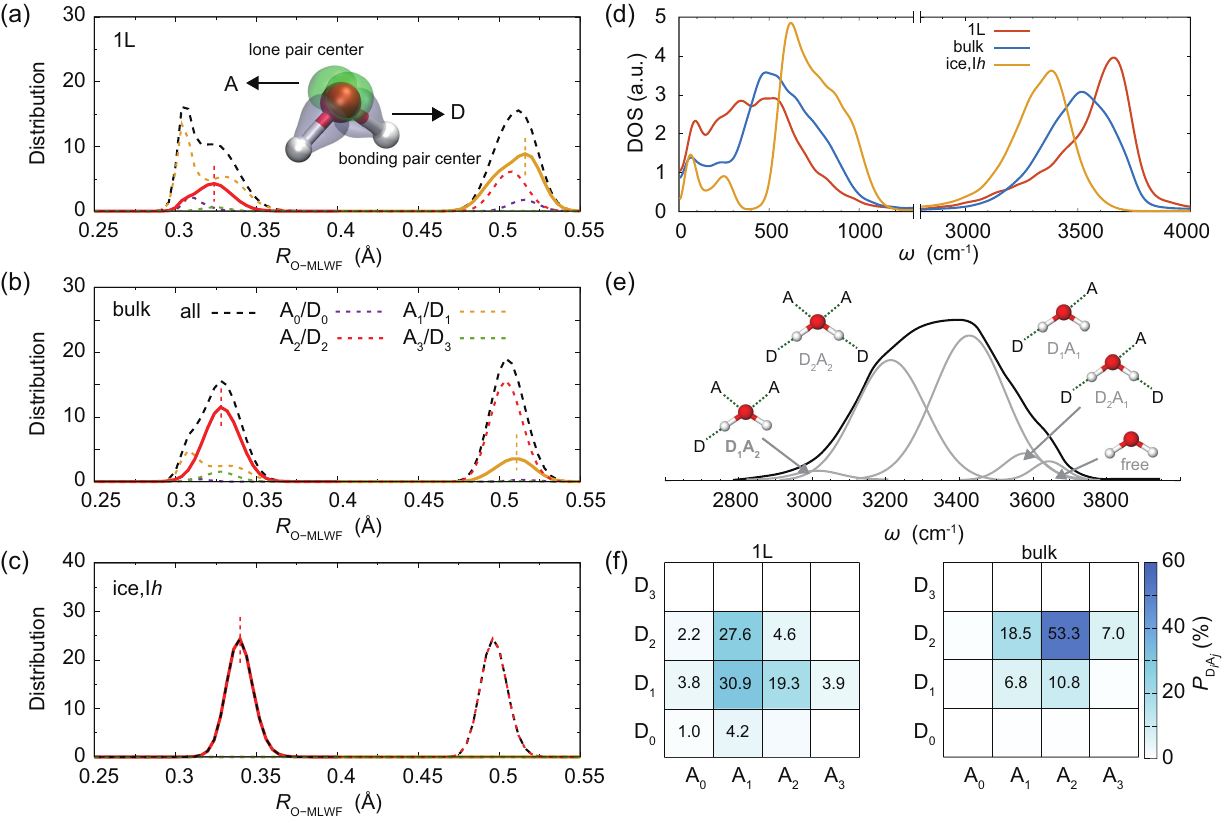} 
\caption{
(a)-(c) Distributions of the centers of maximally localized Wannier functions (MLWFs) with respect to the position of oxygen for lone and bonding electron pairs for bulk water, 1L water and ice I\emph{h}. 
The distributions are also decomposited into $n$ acceptors (A) or donors (D).
The results of A$_2$ and D$_1$ are highlighted by solid and thick lines.
The inset is the representative snapshot of the MLWFs of water molecule.
The lone and bonding pair MLWFs are colored in green and cyan transparent iso-surface, respectively.
(d) The vibrational density of states (DOS) for bulk water, 1L water and ice I\emph{h}.
Here, for clarity, only the bands for the intermolecular vibration and intramolecular OH stretching are shown. For full bands are shown in Fig. S5.
(e) An illustration of the deconvolution of the OH stretching band into five sub-bands, each corresponding to water molecules with different hydrogen-bonding states defined by their number of donors and acceptors \cite{sun2013_cpl}.
(f) The HB statistics of bulk and 1L water in percentage ($\%$) of corresponding water molecules that accept or donate $i=0, 1, 2$ or 3 HBs denoted by $\rm {A}_{\textit i}$ and $\rm {D}_{\textit i}$, respectively \cite{zhou2024_jctca}. 
The background color indicates the percentage ($P({\rm {D}}_i{\rm A}_j)$). The value of $P({\rm {D}}_i{\rm A}_j)$ larger than 1$\%$ is indicated by the number for visualization of the qualitative differences and similarities.
} 
\label{fig3}
\end{figure*}

\textit{\textbf{The lone and bonding electron pairs.}}
To probe these effects, we calculate the centres of lone and bonding electron pairs using maximally localized Wannier functions (MLWFs) \cite{marzari2012_rmp}, which provide insight into the electronic structure across different aqueous environments. 
Figs. \ref{fig3}a-c shows the distributions of the MLWF centers relative to the position of O ($R_{\rm O-MLWF}$) for both lone and bonding electron pairs, along with the decomposition of $n$ acceptors or donors.
In 1L water, the bonding electron center ($R_{\rm O-MLWF} \approx 0.5$ {\AA}) is located farther to the O atom than in the other systems. 
This region acts as a proton-donor site, and a larger $R_{\rm O-MLWF}$ indicates weaker localization of electrons around oxygen, reducing control over the proton and thereby facilitating its release. 
In addition, the full width at half maximum (FWHM) of the MLWF distributions follows the order 1L (0.031 {\AA}) $>$ bulk (0.025 {\AA}) $>$ Ice I\emph{h} (0.020 {\AA}). 
A smaller FWHM (as in Ice I\emph{h}) corresponds to stronger electron confinement and thus stronger O–H covalent bonding. 
In contrast, the larger FWHM in 1L water indicates enhanced electron delocalization, promoting intermolecular electron sharing via hydrogen bonds. 
The increased delocalization in 1L water weakens the covalent O–H bond, making it more prone to stretching and breaking, thereby favoring proton release.
For lone-pair electrons ($R_{\rm O-MLWF} \approx 0.35$ {\AA}), 1L water are located closer to the oxygen atom compared to other systems. 
This reduces the effective electronegativity of oxygen and weakens the directionality of HB. Consequently, the contribution of lone pairs to HB strength follows the order Ice I\emph{h} $>$ bulk $>$ 1L. 
Similarly, distribution of bonding electron, which affects the directionality of hydrogen bonding, exhibits the same trend.
We further performed Mulliken population analysis to quantify the electronic charge distribution on H and O atoms ($\delta_{\rm H}$ and $\delta_{\rm O}$). 
We define $\delta = 2\delta_{\rm H} - \delta_{\rm O}$ as a measure of oxygen’s effective electronegativity and HB directionality (see Fig. S6). 
The order of $\delta$ is Ice I\emph{h} $>$ bulk $>$ 1L, consistent with the analysis of MLWFs.
Our verification calculations using the meta-GGA functional confirm that the conclusions are robust (Fig. S7), reflecting it is the intrinsic properties of nanoconfined water.

\textit{\textbf{The intermolecular and intramolecular vibration.}}
To complement the electronic-structure analysis, we examined the vibrational density of states (DOS) of water to probe both inter- and intramolecular modes.
As shown in Fig. \ref{fig2}d, the 1L water system exhibits an enhanced population of low-frequency intermolecular vibrations ($\omega \lesssim 200$ cm$^{-1}$), associated with collective hydrogen-bond (HB) motions, including bending ($\omega \approx 50$ cm$^{-1}$) and stretching ($\omega \approx 180$ cm$^{-1}$) \cite{zhou2024_jctca}.
These pronounced intermolecular vibrations indicate stronger HB fluctuations and higher HB configuration entropy (see Table S1, see the definition of entropy in \textbf{Methods}), both of which facilitate proton acceptance from broken OH bonds.
For the intramolecular OH stretching band ($\gtrsim 3000$ cm$^{-1}$), 1L water displays an overall blue shift relative to bulk, which seems the OH covalent bond is stiffen and strengthen ($\omega \sim \sqrt{k}$, where $k$ is the bond force constant). 
This shift is expected due to the average HB number in 1L water is reduced (Table S1).
Interestingly, we also observe a pronounced increase in vibrational population near $\omega \approx 3000$ cm$^{-1}$, corresponding to the $\rm{D_1A_2}$ configuration (see the decomposition of the OH stretch mode in Fig. \ref{fig3}e) \cite{sun2013_cpl}, which is consistent with HB statistics that $P({\rm {D}}_1{\rm A}_2)$ is increased for 1L water (Fig. \ref{fig3}d).
The OH bond with lower $\omega$ is prone to have longer bond length \cite{schmidt2007_jpca}, which mean weak bond strength.
The fracture of bond preferentially occurs at the weakest OH bonds which has low $k$ and low $\omega$.
Thus, although the average OH bond stiffness in 1L water is higher, the relative population (the water with ${\rm {D}}_1{\rm A}_2$ configurations) of weakened bonds increases.
Together, the coexistence of weakened OH bonds and amplified HB fluctuations in 1L water suggests a lower barrier for the initial bond-breaking step, consistent with the results of MLP-MD simulations.

\begin{figure*}[tbhp]
\centering
\includegraphics[width=0.90\textwidth]{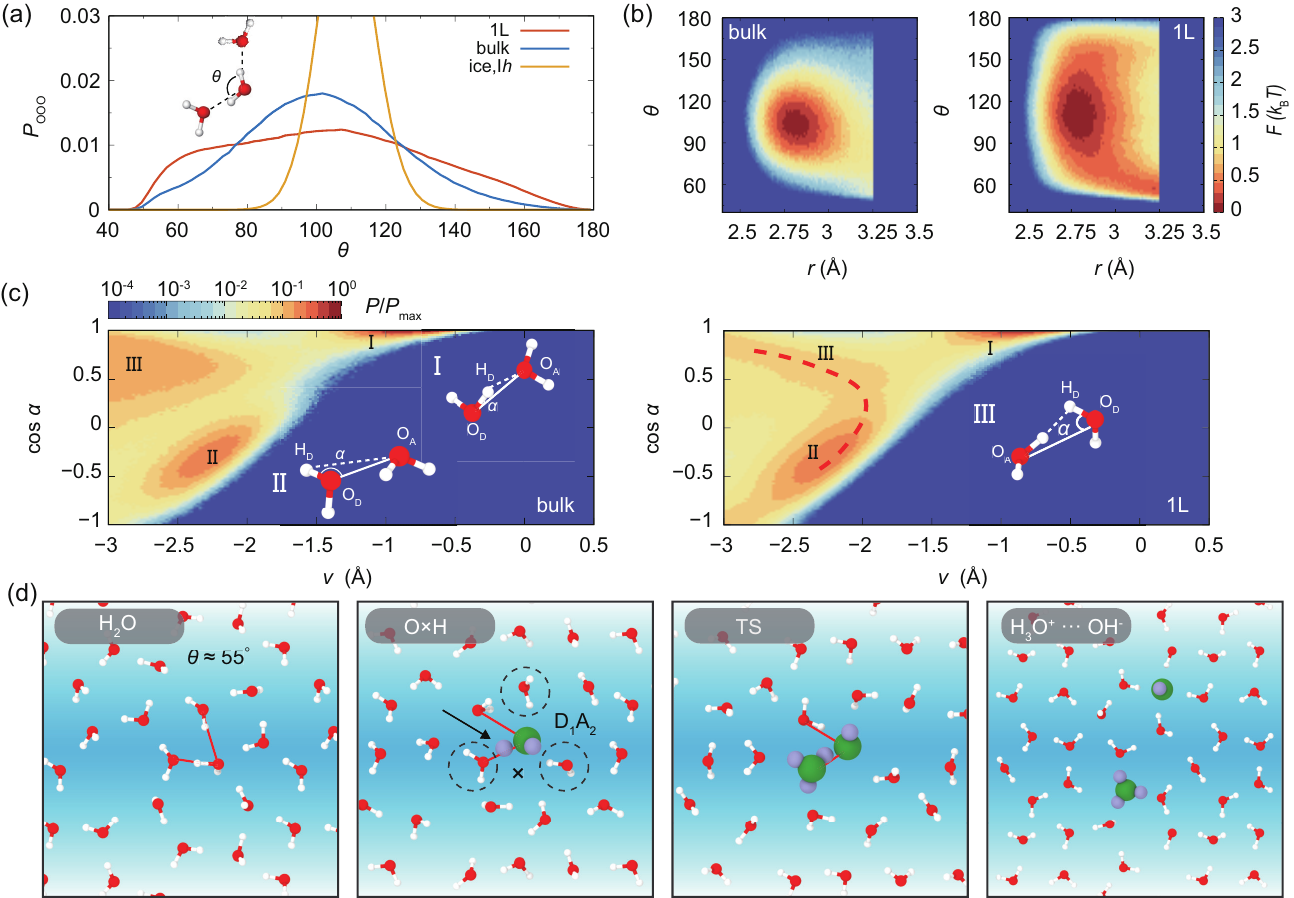} 
\caption{
(a) The bond angular distribution of the triples $P_{\rm {OOO}}(\theta)$. 
The inset is the illustration of $\theta$.
(b) Free energy landscape $F(r, \theta)$ as a function of $\theta$ and the O–O distance $r$ for bulk and 1L water. 
(c) The joint probability distributions of $P(v, {\cos}\,\alpha)$. 
The insets are the schematic representation of the proton-transfer coordinate.
Here $v = d_{\rm {O_{D}H_{D}}} - d_{\rm{O_{A}H_{D}}}$, $\alpha$ is the angle of $\angle$O$_{\rm A}$O$_{\rm D}$H$_{\rm D}$.
Three regions of local minima are indicated by Roman numerals.
The red dash line means the transition path between configuration \uppercase\expandafter{\romannumeral 2} and \uppercase\expandafter{\romannumeral 3}.
(d) Schematic representation of the water auto-dissociation process, including neutral water molecules (H$_2$O), the O···H interaction, the transition state (TS), and the resulting H$_3$O$^+$···OH$^-$ species. The angle indicated by the red lines represents 55$^\circ$ motif. The configuration of ${\rm {D}}_1{\rm A}_2$ is highlighted on the figure.
} 
\label{fig4}
\end{figure*}

\textit{\textbf{The role of interstitial, weak/non- hydrogen-bonded water molecules.}}
Three-body correlations in water can be quantified by the bond angle distribution $P_{\rm OOO}(\theta)$, where $\theta$ is the angle formed by a central oxygen atom and two of its nearest-neighbor oxygen atoms, provides structural insight into hydrogen-bonding environments in water (Fig. \ref{fig2}c) \cite{zhou2024_jctca,chen2017_pnas}.
In bulk water and ice I\emph{h}, the tetrahedral coordination corresponds to $\theta$=109.5$^\circ$, which reflects a well-structured HB network with high stability and directionality.
However, the peak around $\theta$ $\approx$ 55$^\circ$ is associated with disrupted hydrogen bonds or the presence of interstitial, weak hydrogen-bonded water molecules (not like ideal HB with $\theta$=109.5$^\circ$) \cite{chen2017_pnas}.
This triple of 55$^\circ$ means the presence of water molecules occupying interstitial voids within the tetrahedral 
network. 
It indicates partial or complete breakage of the HB network around a central water molecule, indicating the undercoordination and enhanced local disorder.
The free energy profiles of $F(r, \theta)$ are calculated as shown in Fig. \ref{fig2}d, where $r$ is the distance between neighboring oxygen atoms in the triplet.
The increased probability of $\theta \approx$ 55$^\circ$ configuration in the 1L system compared to bulk water and ice I\emph{h} as seen in panel Fig. \ref{fig4}b, along with the lower free energy near $\theta$ $\approx$ 55$^\circ$ and $r=3.25$ {\AA}, indicates greater structural disorder of HB and easy partition of water from interstitial voids, which facilitate the hydration of broken water molecules.

The results of $P_{\rm OOO}(\theta)$ and $F(\theta,r)$ can not very effective in describing fluctuations of the proton. Therefore, we use the proton-transfer coordinate (see the illustration in the inset of Fig. \ref{fig4}c) \cite{ceriotti2013_pnas}, here $v = d_{\rm {O_{D}H_{D}}} - d_{\rm{O_{A}H_{D}}}$, $\alpha$ is the angle of $\angle$O$_{\rm A}$–O$_{\rm D}$–H$_{\rm D}$. 
Three regions of local minima can be found. The configuration of \uppercase\expandafter{\romannumeral 1} ($\cos \alpha \approx 1$, $v \approx -1$ {\AA}) and \uppercase\expandafter{\romannumeral 2} ($\cos \alpha \approx -0.8$, $v \approx -2.5$ {\AA}) refers to ideal HB configuration in first neighbours of the tagged molecule (the water with O$_{\rm D}$). 
While configuration \uppercase\expandafter{\romannumeral 3} with $\cos \alpha \approx 0.5 $ ($\alpha \approx$ 60$^\circ$) is the broken HBs and the interstitial, non–H-bonded water or the non-ideal and weak hydrogen-bonded configurations in the farther neighbours (aligning with the the 55$^\circ$ motif discussed in last paragraph) \cite{ceriotti2013_pnas}.
It can be observed that the 1L system exhibit a continuous pathway with low barrier exists between configuration \uppercase\expandafter{\romannumeral 2} and \uppercase\expandafter{\romannumeral 3}, which corresponds to the transition from the broken HBs/non–H-bonded water to  ideal HB configuration.
We find the dissociation event are strongly relative to the 55$^\circ$ motif (see Fig. \ref{fig2}d) that happens in the triple with 55$^\circ$ (see Fig. \ref{fig4}e).
This motif facilitates the partition of water from interstitial voids by intermolecular motion of HB bend (that is the rotation of water, with pronounced intermolecular vibrations as shown in Fig. \ref{fig3}d) and assist the the hydration of broken water molecules, which reducing the energy barrier of second step for water dissociation.
Here, we also find that when NQE are included, the probability of $v>0$ increases (Fig. S8). This indicates the presence of transient autoprotolysis and suggests that NQE facilitate dissociation.

\section{Discussion}

\begin{figure}[tbhp]
\centering
\includegraphics[width=0.90\textwidth]{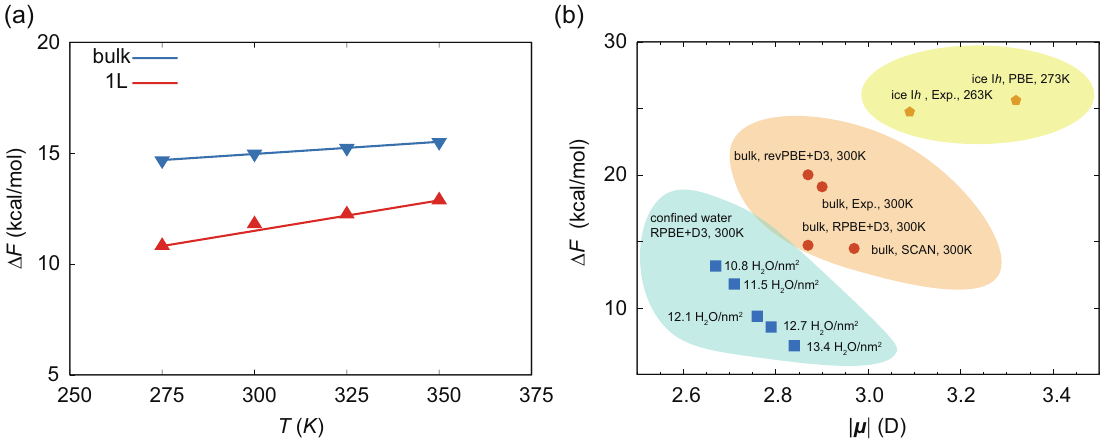} 
\caption{
(a) The dissociation barrier ($\Delta F$) at different temperature.
(b) The relation between $\Delta F$ and dipole moment of water molecule ($|\boldsymbol{\mu}|$).
Results from other methods and experimental data are included for comparison (see Table S2 for a summary).
} 
\label{fig5}
\end{figure}

\textbf{Discussions about in-plane dielectric constant.}
According to experimental observations, the $\Delta F$ for water autoionization is empirically correlated with the $\varepsilon$, with higher $\varepsilon$ leading to lower $\Delta F$ (Fig. S1) \cite{munoz2017_prl}.
This relationship arises because a high dielectric constant enhances charge screening, thereby stabilizing dissociated ions.
Our results therefore imply that $\varepsilon_{\parallel} > \varepsilon_{\rm bulk}$ in 1L water, consistent with recent experimental measurements \cite{wang2024_arxiv}, despite the fact that the out-of-plane dielectric response is extremely weak ($\varepsilon_{\perp} \approx 2$) \cite{geim2018_science}.
Indeed, previous simulations have repeatedly reported anomalously large $\varepsilon_{\parallel}$ ($\varepsilon_{\parallel} \gtrsim 100$) for confined water layers adjacent to channel walls \cite{munoz2017_prl,motevaselian2020_jpcl,bonthuis2012_langmuir,papadopoulou2021_acsnano}.
Such behaviour has been described as \emph{“super permittivity”} \cite{motevaselian2020_jpcl}.
A high $\varepsilon_{\parallel}$ implies strengthened ion hydration ($\sim 1 - 1/\varepsilon$) and weakened ion–ion interactions ($\sim 1/\varepsilon$), reflecting markedly stronger charge screening.
To further probe this, we computed $\Delta F$ as a function of temperature (Fig. \ref{fig5}a) and estimated the entropy change from its slope ($\Delta S = -\partial F/\partial T$) \cite{advincula2025_acsnano}.
We find $\Delta S_{\rm 1L} > \Delta S_{\rm bulk}$, indicating stronger hydration of H$_3$O$^+$ and OH$^-$ under monolayer confinement.
In addition, AIMD simulations of Na$^+$ and Cl$^-$ (see details in \textbf{Methods}) reveal that their hydration shells are more ordered in 1L water than in bulk, with reduced structural entropy ($S^{\rm str}_{\rm 1L} < S^{\rm str}_{\rm bulk}$, Fig. S9), further supporting the view of enhanced hydration.
These findings are consistent with recent MLP-MD work by A. Michaelides and co-workers, who reported that the free-energy level of Na$^+$–Cl$^-$ CIP is much higher than that of separated ions under 1L confinement \cite{fong2024_nanolett}, demonstrating that ion–ion attraction is suppressed while ion hydration dominates.
Together, these results establish that $\varepsilon_{\parallel}$ is strongly amplified in 1L confined water, profoundly reshaping ion hydration and ion–ion interactions.

Here, we summarize the results for $\Delta F$ and the molecular dipole moment ($|\boldsymbol{\mu}|$) across different  systems.
As expected, $|\boldsymbol{\mu}|$ is relatively low in 1L water due to its reduced $n_{\rm HB}$, whereas it is highest in ice I\emph{h} ($n_{\rm HB} \approx 4$).
Overall, we find a general trend that smaller $|\boldsymbol{\mu}|$ correlates with a lower $\Delta F$.
According to the linear-response expression for Ewald summation \cite{zhang2013_jpcl}, the in-plane components of the dielectric tensor are given by
\begin{equation}
\varepsilon_{\alpha \alpha}=1+\frac{4\pi \left( \langle M_{\alpha}^2 \rangle - \langle M_{\alpha} \rangle^2 \right)}{Vk_{\rm B}T\epsilon_0},
\end{equation}
where $M_{\alpha} = \sum_i \boldsymbol{\mu}_{\alpha}^i$ is the total dipole moment of the system and $V$ is the system volume.
For sufficiently large systems, $\langle M \rangle \approx 0$.
Since $\mu_{\rm 1L} < \mu_{\rm bulk}$, maintaining $\varepsilon_{\parallel} > \varepsilon_{\rm bulk}$ requires that dipole fluctuations ($\delta \boldsymbol{M}$) in confinement be larger than those in bulk.
Expressed through the Kirkwood factor, 
\begin{equation}
\varepsilon=1+\frac{N\mu^2}{3k_{\rm B}T\epsilon_0}G_{K}, G_{K}=\frac{\langle \sum_{ij} \bm{\mu_i}   \cdot \bm{\mu_j} \rangle}{N\mu^2}
\end{equation}
The enhanced $\varepsilon_{\parallel}$ means the stronger dipole–dipole correlation in 1L water.
These enhancements in dipole fluctuation and correlation can be rationalized by the fact that geometric confinement enforces quasi-2D alignment, the HB network becomes more cooperative, and soft vibrational modes couple dipoles more strongly. This combination promotes flexible in-plane reorientations and stronger orientational correlations, consistent with the enhanced fluctuations observed in simulations for confined water \cite{zhang2013_jpcl}.

\textbf{Additional remarks.}
We also calculated $\Delta F$ at different water densities (Fig. S10 and Table S3) and found that higher density (refer to the pressure is only about 1 GPa) further facilitates autoionization (Fig. \ref{fig5}b), consistent with previous reports that 1L confined water approaches a superionic state under lateral pressure \cite{kapil2022_nature}. 
Despite their simplicity, protons and hydroxide are among the most fundamental ions in nature \cite{eisenberg2005_water}. 
The discovery that nanoconfinement lowers the barrier for water autoionization provides a missing link between geochemistry and biochemistry. 
The natural confined settings such as clay interlayers, mineral nanopores, or protocell membranes could have acted as \emph{primordial proton factories}, generating reactive species essential for acid–base catalysis \cite{hille2001_ionch,lane2017,kloprogge202_life}. 
These environments may also have established local proton gradients, foreshadowing the central role of chemiosmotic coupling in modern biology \cite{lane2017,hille2001_ionch}. 
Thus, confined water not only reshapes fundamental chemistry but may also have set the stage for life’s first biochemical networks.


In summary, we investigated the confinement effect on water self-dissociation using enhanced-sampling MLP-MD with well-designed global CVs. 
Our results show that extremely confined water is significantly more chemically reactive toward autoionization than bulk water. 
Further analysis reveals that confinement alters the intrinsic structure of water, affecting both the bonding and lone-pair electron localization as well as OH bond vibrations (\emph{intra-molecular}), and the structure and dynamics of the hydrogen-bond network (\emph{inter-molecular})
These changes facilitate OH bond cleavage and the separation and hydration of H$_3$O$^+$ and OH$^-$.
The reduction in $\Delta F$ reflects an anomalously high in-plane dielectric constant, which promotes charge separation and stabilizes dissociated ions. 
Thus, confined monolayer water can be regarded as a \emph{super solvent}, surpassing bulk water’s already extraordinary solvation power. 
We further find that higher density and nuclear quantum effects amplify this tendency, lowering the barrier for proton generation.
Altogether, our findings suggest that life’s origin may have been favored in confined aqueous environments. Confinement provides a natural route to enhanced proton generation, linking geochemical niches to early bioenergetics and echoing the nanoconfined water now exploited in modern catalytic and nanofluidic systems.


\section*{Methods}

\textbf{AIMD and DFT simulations.}
First-principles Born-Oppenheimer MD (BOMD) simulations are performed using the CP2K package \cite{cp2k2014}.
The hybrid Gaussian and plane waves (GPW) scheme is used with a cutoff of 500 Ry \cite{gpw1997}. 
RPBE functional formulated with generalized gradient approximation (GGA) was used to calculate the exchange and correlation energy \cite{RPBE}. 
The van der Waals (vdW) correction of D3 with zero damping proposed by S. Grimme and co-workers is used \cite{grimmeD32010_jcp}.
According to previous work and our calculations, the RPBE+D3 can well describe the structure of liquid water and the dynamical nature of the HB network, approaching the accuracy of CCSD(T) (Figs. S2-S3) \cite{morawietz2016_pnas}.
The triple-$\xi$ quality TZV2P basis set is used for water and the molecularly optimized Gaussian basis sets (mDZVP) are used for C atoms \cite{molopt2007} .
Goedecker-Teter-Hutter (GTH) pseudopotentials are used to treat the core electrons \cite{gth1996}.
The centers of lone and bonding electron pairs are calculated by the MLWFs \cite{marzari2012_rmp}.
The dipole moment of $i-$th water molecule $\boldsymbol{\mu}$ is calculated,
\begin{equation}
\boldsymbol{\mu}=6 \boldsymbol{r}_{\rm O}^i+\boldsymbol{r}_{\rm H_{1}}^i+\boldsymbol{r}_{\rm H_{2}}^i-2\sum^{4}_{c=1}\boldsymbol{r}^{i}_{{\rm W}_{c}}
\end{equation}
where $\boldsymbol{r}_{\rm O}^i$, $\boldsymbol{r}_{\rm H_{1}}^i$ and $\boldsymbol{r}_{\rm H_{2}}^i$ is the position of O and H atoms in the $i$-th water molecules, $\boldsymbol{r}^{i}_{{\rm W}_{c}}$ is the position of the four Wannier centers assigned to the $i$-th water molecules.

AIMD is performed to compare the ion hydration in bulk water and 1L water (see models in Fig. S11).
For the simulation of bulk water system, the simulation box ($12.42 \times 12.42 \times 12.42$ {\AA}) contains 63 water molecules and 1 Na$^+$ or Cl$^-$ as our previous work \cite{zhou2022_jpcb}.
A uniform background charge was employed to neutralize the charge of ions.
For the simulation of 1L water system, the simulation box ($12.298 \times 12.78 \times 13.6$ {\AA}) contains 17 water molecules with 1 Na$^+$ and Cl$^-$ located in each layer of water (Fig. S11c).
For the simulation of ice I\emph{h}, the orthogonal cell ($15.64 \times 13.54 \times 14.72$ {\AA}) is used with 96 water molecule.
The time step is 0.5 fs.  The system is equilibrated in the NVT ensemble using a Nos{\'e}-Hoover thermostat for 100 ps after 10-ps equilibrium by canonical sampling through the velocity rescaling (CSVR) thermostat at 273K.
The trajectory of the last 90 ps is used for analysis.
The number of water molecules in the first hydration shells of ions ($N_{\rm c}$) is determined by using the first minimum in the ion-oxygen radial distribution function (RDF) as a cutoff. 
To quantitatively compare the structural entropy of ions across different aqueous environments, we compute the Gibbs entropy for the rattling motion of water, as proposed in our previous work \cite{zhou2022_jpcb}, 
\begin{equation}
S^{\rm str}=-k_{\rm B}\int P(N_{\rm c})\log(P(N_{\rm c}))d(N_{\rm c}) \nonumber \\
=-k_{\rm B}\sum P(N_{\rm c})\log(P(N_{\rm c})) 
\end{equation}
where $P(N_{\rm c})$ the probability function with $N_{\rm c}$ water in the 1HSs (the first hydrated shell).


\textbf{Definition of collective variables.}
Because the free-energy barrier $\Delta F$ is much larger than $k_{\rm B}T$ ($\Delta F \gg k_{\rm B}T$), the probability for observation of the spontaneous auto-dissociation of water at nanoseconds is impossible. 
Here the enhanced-sampling MD is performed using the advanced collective variables (CVs) based on the Voronoi space partition method \cite{grifoni2019_pnas,zhang2025_Langmuir}. 
The space is tessellated by Voronoi polyhedra centered on the positions of O atoms.
The charge defect of the $i$-th Voronoi center $\delta_i$ can be calculated \cite{grifoni2019_pnas}, with $\delta_i=$ 0, 1, and -1 for H$_2$O, H$_3$O$^+$, and OH$^-$ respectively.
The first CV quantifies the number of auto-dissociated ions ($S_a$),
\begin{equation}
S_{a} = \sum_{i=1}^{N_{\rm{O_w}}} \delta_{i}^2 .
\end{equation}
where $N_{\rm{O_w}}$ is number of O atoms in the system. 
The second CV indicates the distance between the two ions ($S_t$),
\begin{equation}
S_t = -\sum_{i=1}^{N_{\rm{O_w}}} \sum_{j>i}^{N_{\rm{O_w}}} r_{ij} \delta_{i} \delta_{j}.
\end{equation}
where $r_{ij}$ is the distance between two Voronoi centers.
Because the CV of $S_t$ cannot well identify the states between pure water and transition states \cite{zhang2025_Langmuir}, the $S_t$ is converted to its logarithmic variant to amplify the distance in path of $S_t$,
\begin{equation}
    S'_{t} = \begin{cases}
\log({S}_t + \epsilon), & \ 0 \leq {S}_t < 1, \\
{S}_t - 1 + \log(1+\epsilon), & \ {S}_t \geq 1,
\end{cases}
\end{equation}
where $\epsilon = 0.03$ is a regularization parameter. 
The self-ion distance variant $S_t^{'}$ can differentiate between the pure water state ($S_t \approx 0.0$ {\AA}, $S_t’ \approx -3$ {\AA}), the transition state ($S_t \approx 2.9$ {\AA}, $S_t’ \approx 2.0$ {\AA}), and the autoionization state ( $S_t \approx 6.0$ {\AA}, $S_t’ \approx 4.8$ {\AA}).
The breaking of the OH covalent bond was defined to occur when the O-H distance reached the position of the first minimum in the O-H RDF, $g_{\rm OH}(r)$ (see Fig. S12), corresponding to values of $S_a \approx 0.7$ {\AA}, $S_t’ \approx 0.2$ {\AA}.

\textbf{The train of MLPs.}
The MLPs used in this work were developed with the neuroevolution potential (NEP) framework implemented in the GPUMD package \cite{fan2021_prb,fan2022_jcp}.
The training database consists of two parts, including pure water configurations sampled from AIMD trajectories over a broad temperature range (300–800 K), and ion+water configurations generated by applying targeted restraints during AIMD.
To represent the OH bond cleavage, the coordination number $n_{\rm H}$ for the target O$^*$ is defined,
\begin{equation}
n_{\rm H}=\sum_i^{N_{\rm{H_w}}} \frac{1-(r_i/R_0)^{p}}{1-(r_i/R_0)^{q}}
\end{equation}
where where $r_i$ is the O$^*$H distance, the cutoff distance $R_0 = 1.38$ {\AA}, $p=12$, $q=24$ \cite{joutsuka2022_jpcb}.
The value of $n_{\rm H}$ is constrained to increase (decrease) from $\approx 2$ to $\approx 3$ (1), driving the gradual transformation of a neutral water molecule into OH$^-$ (H$_3$O$^+$) under a harmonic potential (spring constant 2000 kj/mol).
Similarly, configurations at different $S_t$ values are sampled using a harmonic constraint (spring constant 2000 kj/mol), ensuring adequate sampling of solvation shells and ion–ion interactions.
The configurations of PIMD simulation at 300K is also added. 
Here, incorporating high-temperature structures enables the comprehensive capture of configurations featuring significant elongation of OH bonds and distortion of $\angle$HOH. 
More details of train sets can refer to Tables S4-S5.
The training sets are randomly selected from these trajectories of AIMD for the train of first NEP model.
Then the training sets are selected based on the farthest-point sampling by comparing the distance of structures in 2D principal component space for the final NEP model.
The training loops for MLPs are more than 10$^6$ that ensure the root mean squared error (RMSE) on the total energy and atomic force per molecule are $\lesssim$ 1.5 meV/atom and $\lesssim$ 80 meV/{\AA}, respectively (see Fig. S13).
To confirm the reliability of the training MLP, we compare the results of RDFs and HB statistics (Figs. S14-S15) that the MLP can well reproduce the results of AIMD and PIMD, which means the MLP is reliable.

\textbf{The details of MLP-MD simulations.}
Molecular dynamics (MD) simulations were performed using the LAMMPS package \cite{lammps1995}. Enhanced sampling was carried out with the PLUMED plugin via the on-the-fly probability enhanced sampling (OPES) method to accelerate water autoionization events \cite{tribello2014,invernizzi2020_opes}.
The bulk system contained 64 water molecules in a cubic box of 12.42 {\AA}, while the 1L system comprised 18 molecules per layer in a $12.29 \times 12.78 \times 13.60$ {\AA} cell (Fig. S16). 
Previous studies have shown that systems containing 64 water molecules are sufficient to achieve converged $\Delta F$ \cite{liu2023_prl}.
Our calculations further confirm that this system size yields well-converged free-energy surfaces, ensuring the robustness of our conclusions (Fig. S17).
Periodic boundary conditions were applied in all directions.
All simulations were carried out in the canonical (NVT) ensemble at 300 K using a Nos{\'e}–Hoover chain thermostat, with a 0.5 fs time step and production runs of 10 ns.
To model water autoionization, OPES was applied using the collective variables $S_a$ and $S’_{t}$.
All oxygen and hydrogen atoms were globally labelled to prevent bias toward specific molecules and to capture the intrinsic dynamics of the system. 
The bias potential was updated every 500 steps, striking a balance between obtaining smooth probability estimates and avoiding noisy fluctuations from excessively frequent updates.

\textbf{The structure and dynamical analysis of water and HBs.}
The geometry-based criterion is used to analyze the HB network, including the distance between the oxygen of both molecules smaller than 3.6 {\AA} and the angle defined within the dimer geometry ($\angle${O-O-H}) smaller than 30$^\circ$ \cite{luzar1996_nature}.
For each water molecule, the numbers of donors (D) and acceptors (A) are counted, and the joint probability distribution $P({\rm {A}}_i{\rm D}_j)$ is obtained \cite{zhou2024_jctca}. 
Here $\rm {A}_{\textit i}$ or $\rm {D}_{\textit i}$ means accept or donate $i=0, 1, 2$ or 3 HBs. 
The number of HB for each water molecule can be expressed as, $n_{\rm HB}= \langle i+j \rangle$.
The structural entropy of HB can be defined as, 
\begin{equation}
S_{\rm HB}=-k_{\rm B}\sum P({\rm {A}}_i{\rm D}_j)\log(P({\rm {A}}_i{\rm D}_j)) 
\end{equation}
The vibration density of states (DOS) is calculated as the Fourier transform of the velocity autocorrelation function (VACF),
\begin{equation}
{\rm DOS}(\omega)=\int\langle\mathbf{v}(0)\cdot\mathbf{v}(t)\rangle e^{-i\omega t}dt
\end{equation}
where $\mathbf{v}(t)$ is the atomic velocities.

\textbf{PIMD simulations.}
The PIMD simulation was carried out by i-PI package \cite{kapil2019cpc} interfaced with  LAMMPS, where i-PI collects the atomic forces and biased forces calculated from NEP to propagate the nuclear motion. 
A generalized Langevin equation thermostat (PIGLET) was applied to the path integral sampling using 12 beads per atom \cite{ceriotti2012prl}. 
The PIMD with 12 beads can get well converged results (see Fig. S18).
The enhanced sampling PIMD simulation was also performed with the PLUMED plugin employing the OPES methods.

\textbf{The calculation of pKw.}
The dissociation constant, pK${\rm w}$, can be evaluated from the free-energy profile of $F(S_t)$ \cite{calegari2023_pnas},
\begin{equation}
{\rm pK_w} = 2\log\left[c_0 \int_0^{S_c} e^{-\beta \Delta F(S_t)} 4\pi S_t^2, dS_t \right],
\end{equation}
where $c_0 = 1/1660$ Å$^3$ (1 M) is the standard concentration and $S_c$ denotes the dividing distance between H$_3$O$^+$ and OH$^-$.
The entropic contribution is accounted for by the $4\pi S_t^2$ factor in the integral \cite{calegari2023_pnas}.
For $S_t > 6$ {\AA}, the reaction products are well separated and $F(S_t)$ converges to a constant value and thus we set $S_c = 6$ {\AA} in this work.
Following previous studies, pK${\rm w}$ can also be directly estimated from the free-energy barrier as ${\rm pK_w} = \Delta F / (RT \ln 10)$ \cite{joutsuka2022_jpcb}.
When either $\Delta F$ or pK${\rm w}$ values are unavailable (Table \ref{table1}), the missing quantity is computed using this relationship.

\section*{Author Contributions}
K.Z. conceived and directed the research. All authors analyzed data and wrote the manuscript.

\section*{Acknowledgements}

The authors acknowledge the financial support of
 the National Natural Science Foundation of China (No. 12572126), 
 the Natural Science Foundation of the Higher Education Institutions of Jiangsu Province, China (No. 25KJB120008). 
The work was carried out at the National Supercomputer Center in Tianjin, and the calculations were performed on TianHe-1(A).


\section*{Data availability}
Original data for machine learning potential for the study will be available on the Materials Cloud platform ((https://doi.org/x/materialscloud:x). 
The input files for PLUMED will be available on the PLUMED-NEST (plumID:X.X).
The data within this paper and other findings of this study are available from the corresponding authors upon reasonable request.

\section*{Code availability}
The codes and input scripts used within this work are available from the corresponding authors upon reasonable request.

\section*{Competing Interests}
The authors declare no competing interests.


\bibliographystyle{naturemag}


\end{document}